\documentstyle[preprint,aps]{revtex}
\newcommand{\beq}{\begin{eqnarray}}
\newcommand{\eeq}{\end{eqnarray}}
\newcommand{\nnb}{\nonumber}

\begin{document}
\draft  
\preprint{\vbox{\baselineskip16pt
\hbox{AS-ITP-2000-001}
\hbox{SNUTP 00-006}
\hbox{hep-ph/0006003}
}}
\title{1/${\sf m_Q}$ Corrections to the Heavy-to-Light-Vector Transitions 
in the HQET}
\tightenlines
\author{Chao-Shang Huang$^a$\footnote{Email address:~csh@itp.ac.cn}, 
Chun Liu$^{a,b}$\footnote{Email address:~liuc@itp.ac.cn}, and 
Chi-Tau Yan$^a$\footnote{Email address:~cyan@itp.ac.cn}} 
\address{
$^a$Institute of Theoretical Physics, Chinese Academy of Sciences,\\ 
P.O. Box 2735, Beijing 100080, China\\
$^b$Department of Physics and Center for Theoretical Physics,\\
    Seoul National University, Seoul 151-742, Korea}
\date{\sl \today}
\maketitle

\begin{abstract}
Within the HQET, the heavy to light vector meson transitions are 
systematically analyzed to the order of $1/m_Q$.  Besides the four universal 
functions at the leading order, there are twenty-two independent universal form 
factors at the order of $1/m_Q$.  Both the semileptonic decay $B\to\rho$ which 
is relevant to the $|V_{ub}|$ extraction, and the penguin induced decay 
$B\to K^*$ which is important to new physics discovering, depend on these form 
factors.  Phenomenological implications are discussed.  

\pacs{PACS number(s): 12.39.Hg, 13.20.He }
\end{abstract}

There are two main reasons to study the heavy meson to light meson weak 
transitions.  To extract the Cabibbo--Kobayashi--Maskawa matrix element 
$V_{ub}$ precisely which has important implications for CP violation, 
the exclusive $B\to\rho (\pi)\,\ell\,\nu$ decays are suitable channels 
\footnote{An alternative way for the $|V_{ub}|$ extraction is from the 
inclusive $B\to X_u$ decays.}.
Another reason is to investigate the rare B exclusive 
decays induced by penguin diagrams which are important for testing the 
standard model and for discovering new physics.  They are the processes 
$B\to K^*\gamma$ and $B\to K^{(*)}l^+l^-$, respectively.  Large number of 
samples of the B to light meson processes produced in the current 
experiments~\cite{CLEO} and to be produced in the near future B factories will 
make precise measurements available.  Thus the main task is to reduce the 
theoretical uncertainties in the calculations of the hadronic matrix elements.

In this paper, we focus on the B to light vector decays.  The matrix elements 
responsible for the decay $H\to V\,\ell\,\nu$ can be parameterized in terms of
four invariant form factors, which are conventionally defined as
\beq
\langle V(p,\epsilon) |\,\bar q\,\gamma_\mu\, Q\,| H(P)\rangle
&=& i\,g(q^2)\, \varepsilon_{\mu\nu\lambda\sigma}\, \epsilon^{*\nu}\,
  (P+p)^\lambda\, (P-p)^\sigma \,, \nonumber\\*
\langle V(p,\epsilon) |\,\bar q\,\gamma_\mu\gamma_5\, Q\,| H(P)\rangle
&=& f(q^2)\,\epsilon^*_\mu
  + a_+(q^2)\,(\epsilon^*\cdot P)\,(P+p)_\mu 
  + a_-(q^2)\,(\epsilon^*\cdot P)\,(P-p)_\mu \,,\label{1}
\eeq
where $q^2=(P-p)^2$.  The matrix elements for the decays $B\to K^*\gamma$ and 
$B\to K^*l^+l^-$ are parameterized by the following three invariant form 
factors,
\begin{eqnarray}
\langle V(p,\epsilon) |\,\bar q\,\sigma_{\mu\nu}\, Q\,| H(P)\rangle
&=& g_+\, \varepsilon_{\mu\nu\lambda\sigma}\, \epsilon^{*\lambda}\,
  (P+p)^\sigma + g_-\, \varepsilon_{\mu\nu\lambda\sigma}\,
  \epsilon^{*\lambda}\, (P-p)^\sigma  \nonumber\\*
&+& h\, \varepsilon_{\mu\nu\lambda\sigma}\, (P+p)^\lambda\,
  (P-p)^\sigma\, (\epsilon^* \cdot P) \,,\nnb\\
  \langle V(p,\epsilon) |\,\bar q\,\sigma_{\mu\nu}\gamma_5\, Q\,| H(P)\rangle
&=& i\,g_+\,[\epsilon^*_\nu\,(P+p)_\mu-\epsilon^*_\mu\,(P+p)_\nu]
  \nonumber\\*
&+& i\,g_-\,[\epsilon^*_\nu\,(P-p)_\mu-\epsilon^*_\mu\,(P-p)_\nu]
  \nonumber\\*
&+& i\,h\, [(P+p)_\nu\,(P-p)_\mu-(P+p)_\mu\,(P-p)_\nu]\,
  (\epsilon^*\cdot P) \,. \label{t2}
\end{eqnarray}
Here the second relation is obtained from the first one using
$\sigma^{\mu\nu}=\frac{i}2\,\varepsilon^{\mu\nu\lambda\sigma}\,
\sigma_{\lambda\sigma}\,\gamma_5$.

We use the heavy quark effective theory (HQET)~\cite{2} to study these form 
factors to the order of $1/m_Q$.  The HQET provides a clear physical 
description for the hadrons containing a single heavy quark.  It has been 
successfully applied to the analysis of the $B\to D^{(*)}$ decays.  It has 
been also used for the heavy to light meson exclusive weak decays~\cite{3}.  
In this latter case, the form factors have no normalization point.  The heavy 
quark symmetry (HQS) does not simplify the analysis significantly.  
Nevertheless, the relations between various heavy to light 
meson transitions can be found by the HQS.  
At the order of $1/m_Q$ many form factors are introduced.  However, they are 
universal for all the heavy to light transitions.  The systematic 
nature of the heavy quark expansion means uncertainties are easier to identify 
and estimate.  Furthermore the analysis 
based on HQET is model-independent.  
It is therefore meaningful to consider the $1/m_Q$ corrections to the 
heavy to light meson transitions, in addition to the leading 
order results.  They are also practically important for the analysis of the 
$D\to\rho~,K^*$ weak decays.  For the heavy to light pseudo-scalar weak 
decays, the $1/m_Q$ corrections have been considered in Ref.~\cite{5}.  We 
will calculate the $1/m_Q$ corrections to the heavy to light vector meson 
transitions.  

Let us make a brief review of the HQET.  In the heavy quark limit, the 
velocity of the heavy quark $Q$, $v$, is a well defined quantity and the heavy 
quark field can be represented by the velocity-dependent field,
\begin{equation}\label{EF}
  h_v(x) = \exp({i m_Q\,v\!\cdot\! x)\,P_+\,Q(x) \,,}
\end{equation}
where 
$P_{+}={1+\rlap/v\over 2}$. 
The effective Lagrangian is
\begin{equation}\label{Leff}
   {\cal{L}}_{\rm eff}^0 = 
   \bar h_v\,i v\!\cdot\!D\,h_v 
   \,,
\end{equation}
where the gauge-covariant derivative generates the residual momentum.  To the 
$1/m_Q$ order~\cite{2,6}, the relation between $Q$ and $h_v$ is obtained by 
treating $1/m_Q$ as perturbation,
\beq
Q(x) = \exp(- i m_Q\,v\!\cdot\! x)\,(1 + \frac{i\,\rlap/\!D_\perp}%
{2m_Q}) h_v(x)\,,
\eeq
where $D^{\mu}_\perp = D^{\mu} -v^{\mu}\,v\cdot D$, and $h_v$ satisfies 
exactly the equation of motion $i\,v\cdot D\, h_v =0$. 
The effective Lagrangian becomes,
\begin{equation}\label{Leff2}
{\cal{L}}_{\rm eff} = {\cal{L}}_{\rm eff}^0
+ {\frac{1}{2 m_Q}}\,\Big[ O_{\rm kin}
+ \,O_{\rm mag} \Big] + {\cal{O}}(1/m_Q^2) \,,
\end{equation}
where
\begin{equation}\label{Omag}
O_{\rm kin} = \bar h_v\,(i D)^2 h_v \,, \qquad
O_{\rm mag} = {g_s\over 2}\,\bar h_v\,\sigma_{\mu\nu}
G^{\mu\nu} h_v \,.
\end{equation}
$O_{\rm kin}$ describes the kinetic energy of the heavy quark in the hadron,
and $O_{\rm mag}$ the heavy quark chromomagnetic energy.

To study the hadronic matrix elements in the HQET, the form factors are 
considered as functions of the kinematic variable
\begin{equation}
   v\cdot p = {m_H^2 + m_V^2 - q^2\over 2 m_H} \,.
\end{equation}
Accordingly, Eqs. (\ref{1}) and (\ref{t2}) can be re-expressed as  
\beq
\langle V(p,\epsilon) |\,\bar q\,\gamma_\mu\, h_v\,| H(v)\rangle
&=&2i
  \,\tilde g(v\cdot p)\, \varepsilon_{\mu\nu\lambda\sigma}\, \epsilon^{*\nu}\,
  \hat p^\lambda\, v^\sigma \,, \nnb\\*
\langle V(p,\epsilon) |\,\bar q\,\gamma_\mu\gamma_5\, h_v\,| H(v)\rangle
&=& 2
  \left[\tilde f(v\cdot p)\,\epsilon^*_\mu
  + \tilde a_1(v\cdot p)\,(\epsilon^*\cdot v)\,\hat p_\mu 
  + \tilde a_2(v\cdot p)\,(\epsilon^*\cdot v)\,v_\mu\right] \,,
\eeq
and
\begin{eqnarray}
\langle V(p,\epsilon) |\,\bar q\,\sigma_{\mu\nu}\, Q\,| H(v)\rangle
&=& 2
 \left[\tilde g_v\, \varepsilon_{\mu\nu\lambda\sigma}\, \epsilon^{*\lambda}\,
  v^\sigma + \tilde g_p\, \varepsilon_{\mu\nu\lambda\sigma}\,
  \epsilon^{*\lambda}\, \widehat p^\sigma
  + \tilde h\, \varepsilon_{\mu\nu\lambda\sigma}\, v^\lambda\,
  \widehat p^\sigma\, (\epsilon^*\cdot v)\right] \,, \nnb\\
  \langle V(p,\epsilon) |\,\bar q\,\sigma_{\mu\nu}\gamma_5\, Q\,| H(v)\rangle
&=&2 
  \Big\{i\,\tilde g_v'\,[\epsilon^*_\nu\,v_\mu-\epsilon^*_\mu\,v_\nu]
  + i\,\tilde g_p'\,[\epsilon^*_\nu\,\widehat p_\mu-\epsilon^*_\mu\,
  \widehat p_\nu]
  \nonumber\\*
&&\quad + i\,\tilde h'\, [v_\nu\,\widehat p_\mu-v_\mu\,\widehat p_\nu]\,
  (\epsilon^*\cdot v)\Big\} \,,
\end{eqnarray}
where the dimensionless variable is 
\begin{equation}
   \widehat p^\mu = {p^\mu\over v\cdot p} \,,\qquad
   v\cdot\widehat p = 1 \,,
\end{equation}
so that all of the form factors have the same dimension.  It is convenient to 
work in the matrix representation of the hadrons~\cite{7}.  These wave 
functions are only dependent on the HQS and their Lorentz 
transformation properties.  The ground-state pseudo-scalar and vector heavy 
mesons are described by 
\begin{equation} \label{EFST}
   {\cal{M}}(v) = 
   {1+\rlap/v\over 2}\,
   \cases{ -\gamma_5\,, & pseudoscalar meson; \cr
           \rlap/\epsilon\,, & vector meson with polarization
            vector $\epsilon^\mu$ . \cr}
\end{equation}
Based on the symmetry and the Feynman rules of the HQET, one can express the 
hadronic matrix element by evaluating some trace over the above wave 
functions. At the leading order of $1/m_Q$, the matrix element of the relevant 
current $\bar q \Gamma h_v$ can be written as
\begin{equation}\label{LOtrace}
  \langle V(p,\epsilon)|\,\bar q\,\Gamma\,h_v\,|H(v)\rangle
  = - {\rm Tr}\Big\{ \Omega_L (v,\, p)\,\Gamma\,{\cal{M}}(v) \Big\} \,,
\end{equation}
where the matrix $\Omega_L(v,\, p)$ transforms as a Lorentz scalar as 
functions of $v\cdot p$.  And it has linear dependence on the polarization of 
the meson $V$.  Considering ${\cal{M}}(v)\,\rlap/v = -{\cal{M}}(v)$, the 
general form for $\Omega_L$ is 
\begin{equation}
 \Omega_L = 
 L_1\rlap/\epsilon^\ast +
 L_2v\cdot \epsilon^\ast +
 [ L_3\rlap/\epsilon^\ast +
 L_4v\cdot \epsilon^\ast ]\,\rlap/\! \widehat p
\,,
\end{equation}
where the universal functions $L_i$ ($i=1 \sim 4$) depend on the kinematic 
variable 
$v\cdot p$, but not on the heavy quark mass $m_Q$.

The power corrections proportional to $1/m_Q$ result from both the effective
currents and  the effective Lagrangian of the HQET.  We first consider the 
corrections coming from the expansion of the currents.  Weak current of the 
heavy-to-light transition in the effective theory can be expanded as,
\beq \label{current}
\bar q \Gamma Q =\bar q \Gamma
(1 + \frac{i\,\rlap/\!D_\perp} {2m_Q}) h_v(x)\,.
\eeq
In the same manner as shown in leading order, one can find that the matrix 
elements of the operators containing a covariant derivative which acts on the 
heavy quark field have the formal structure,
\begin{eqnarray}
   \quad \langle V(p,\epsilon)|\,\bar q\,\Gamma\,i\, D^\mu h_v\,|M(v)\rangle 
   = - {\rm Tr}\Big\{ \Omega_D^{\mu}(v,\, p) \Gamma\,{\cal{M}}(v) \Big\} \,.
\end{eqnarray}
The matrix $\Omega_D^{\mu}(v,\, p)$ also contains some universal functions 
depending only on the variable $v\cdot p$ , and transforms as a vector.  
The generic structure of $\Omega_D^\mu$ is 
\begin{eqnarray}
   \Omega_D^\mu&=& ( D_1\,v^\mu + D_2\,\widehat p^\mu
    + D_3\,\gamma^\mu)\,\rlap/\epsilon^\ast 
   + ( D_4\,v^\mu + D_5\,\widehat p^\mu
    + D_6\,\gamma^\mu)\,(v\cdot \epsilon^\ast) \nnb \\
  && + 
    ( D_7\,v^\mu + D_8\,\widehat p^\mu
    + D_9\,\gamma^\mu)\,\rlap/\epsilon^\ast \rlap/\!\widehat p 
   + ( D_{10}\,v^\mu + D_{11}\,\widehat p^\mu
    +  D_{12}\,\gamma^\mu)\,(v\cdot \epsilon^\ast) \,\,\rlap/\!\widehat p \nnb \\ 
   && + ( D_{13} + D_{14}\,\,\rlap/\!\widehat p )\,
    \epsilon^{\ast{\mu}}
    \,.
\end{eqnarray}
The functions $D_i$ are functions of $v\cdot p$.  Not all of these fourteen 
universal functions are independent.  Using equation of motion of the heavy 
quark, $i v\!\cdot\!D\,h_v=0$, we can easily obtain
\begin{eqnarray}\label{rel12}
   D_1 + D_2 - D_3 &=& 0\,, \nonumber\\
   D_4 + D_5 - D_6 +D_{13} &=&0 \,, \nnb \\
   D_7 + D_8 - D_9 &=&0\,, \nnb \\
   D_{10} +D_{11} - D_{12} +D_{14} &=&0 \,.
\end{eqnarray}
Furthermore, using the following relations,
\begin{equation}
   i\partial^\mu (\bar q\,\Gamma\,h_v) = \bar q\,\Gamma\,(i D^\mu)\,h_v 
   +i\bar q\,(\overleftarrow{D^\mu})\,\Gamma\,h_v\,,
\end{equation}
and
\begin{equation}\label{totalder}
   \langle V(p, \epsilon)|\,i\partial^\mu(\bar q\,\Gamma\,h_v)\,|H(v)\rangle
   = (\bar\Lambda\,v^\mu - p^\mu)\,
   \langle V(p,\epsilon)|\,\bar q\,\Gamma\,h_v\,|H(v)\rangle \,,
\end{equation}
where $\bar\Lambda=m_M-m_Q$ denotes the finite mass difference between a heavy 
meson and the heavy quark in the infinite quark mass limit, and equation of 
motion for the light quark field, $i\,\rlap/\!D\,q=0$, we can obtain
\beq
{\rm Tr}\,\Big\{ \Omega^\mu_D(v,p)\,\gamma_\mu \,\Gamma'\,{\cal M}(v) \Big\}\,=
\,(\bar\Lambda v^\mu- p^\mu)\,{\rm Tr}\,\Big\{ \Omega_L(v,p)\,\gamma_\mu\,
\Gamma'\,
{\cal M}(v) \Big\}\,,
\eeq
where we have substituted $\Gamma$ by $\gamma_\mu \,\Gamma'$.  It yields,
\begin{eqnarray}\label{rel34}
   D_1-2D_3+2D_7+\widehat p^2D_8+D_{13} &=& -\bar\Lambda L_1 - (2\bar
   \Lambda -v\cdot p \,\widehat p^2) L_3 \,,\nnb \\
   2D_2-D_4+4D_6+2D_{10}+\widehat p^2 D_{11} &=& \bar \Lambda (L_2-2L_1)
    - (2\bar\Lambda - v\cdot p\,\widehat p^2)\,L_4 \,, \nnb \\
    D_2-D_7-D_{14} &=& \bar\Lambda L_3 + v\cdot p \,L_1\,, \nnb \\
    D_5-2D_7+D_{10}-2D_{12} &=& \bar\Lambda (2 L_3 - L_4)+v\cdot p L_2 \,.
\end{eqnarray}
The relations, Eqs.~(\ref{rel12}) and (\ref{rel34}), imply that only six of 
the fourteen universal functions are independent.  Note that the light quarks 
have been taken to be massless.  This reduces the number of HQET operators 
appearing in the expansion of QCD currents.  

The corrections to the effective states should be included.    
The $1/m_Q$ terms in the effective 
lagrangian are treated as perturbation, $h_v$ is still defined by Eq. 
(\ref{Leff}) at the subleading order of the heavy quark expansion.  Therefore 
the effective states of Eq.~(\ref{EFST}) are not the eigenstates of the 
operators $O_{\rm kin}$ and $O_{\rm mag}$.  The corrections to the effective 
states can be accounted for by including 
time-ordered products in which $O_{\rm kin}$ or $O_{\rm mag}$ is inserted into 
matrix elements of the leading-order currents.  By using the Feynman rules in 
HQET, one can obtain
\begin{eqnarray}
   &&\langle V (p,\epsilon)|\,i\int{\rm d}y\,T\Big\{ \bar q\,\Gamma\,h_v(0),
   O_{\rm kin}(y) + O_{\rm mag}(y)\Big\}\,|H(v)\rangle \nonumber\\
   &&\quad = - {\rm Tr}\bigg\{ \Omega_K(v,\, p)\,\Gamma\,{\cal{M}}(v)
   \,+\, \Omega_G^{\alpha\beta} (v,\, p)\,\Gamma\,{1+\rlap/v\over 2}\,
   \sigma_{\alpha\beta}\,{\cal{M}}(v) \bigg\} \,,
\end{eqnarray}
where the properties of matrix $\Omega_K(v,\, p)$ are very similar to matrix
$\Omega_L$, and the matrix $\Omega_G^{\alpha\beta}(v,\, p)$ also has the 
similar properties except it must transform as a tensor.  They 
can be described in terms of sixteen additional universal functions 
$S_i(v\cdot p)$ as follows:
\begin{eqnarray}
  \Omega_K&=& S_1\,\rlap/\epsilon^\ast +
  S_2(v\cdot \epsilon^\ast)\, +
  [ \,S_3\,\rlap/\epsilon^\ast +
  S_4(v\cdot \epsilon^\ast)\, ]\,\,\rlap/\! \widehat p
  \,, \nonumber \\
  \Omega_G^{\alpha\beta}&=& 
  ( i S_5 \widehat p^\alpha \gamma^\beta + \,S_6 \sigma^{\alpha\beta} )
   \,\rlap/\epsilon^\ast \,\,+ 
   ( i S_7 \widehat p^\alpha \gamma^\beta + \,S_8 \sigma^{\alpha\beta} )
   (v\cdot \epsilon^\ast) 
    + ( i S_9 \widehat p^\alpha \gamma^\beta + \,S_{10} \sigma^{\alpha\beta} )
   \,\rlap/\epsilon^\ast\,\,\rlap/\!\widehat p  \nnb \\
   &&+ ( i S_{11} \widehat p^\alpha \gamma^\beta + \,
   S_{12} \sigma^{\alpha\beta} )
   \,\,\rlap/\!\widehat p \,(v\cdot \epsilon^\ast)
   + (i S_{13} \gamma^\alpha 
   \epsilon^{\ast\beta } + \, i S_{14} \gamma^\alpha \epsilon^{\ast\beta}\,
   \,\rlap/\!\widehat p )\, \nnb \\
   &&+ (iS_{15}\epsilon^{\ast\alpha}\widehat{p}^\beta 
  + iS_{16}\epsilon^{\ast\alpha}\widehat{p}^\beta\,\rlap/\!\widehat p )\,.
\end{eqnarray}
To get the independent universal functions in $\Omega_G^{\alpha\beta}$,
let us consider the following matrix element of the time-ordered 
operator products~\cite{dgh},
\begin{eqnarray}
\langle V (p,\epsilon)|\,i\int{\rm d}y\,T\Big\{ \bar q\,\Gamma\,h_v(0),
   \frac{i}{2}g_s \bar h_v\Gamma_1\,G^{\alpha\beta} h_v \Big\}\,|H(v)\rangle 
  = - {\rm Tr}\bigg\{ 
   \, \bar\Omega_G^{\alpha\beta} (v,\, p)\,\Gamma\,{1+\rlap/v\over 2}\,
   \Gamma_1\,{\cal{M}}(v) \bigg\} \,,
\end{eqnarray}
where $\Gamma_1$ is any fixed Dirac matrix, and 
\beq
 \bar \Omega_G^{\alpha\beta}&=&\frac{1}{2}\Big[ 
   ( i S_5 \widehat p^{[\alpha} \gamma^{\beta]} + 2S_6 \sigma^{\alpha\beta} )
   \,\rlap/\epsilon^\ast \,+ 
   ( i S_7 \widehat p^{[\alpha} \gamma^{\beta]} + 2S_8 \sigma^{\alpha\beta} )
   (v\cdot \epsilon^\ast) 
    + ( i S_9 \widehat p^{[\alpha} \gamma^{\beta]} +2S_{10} \sigma^{\alpha\beta} )
   \,\rlap/\epsilon^\ast\,\,\rlap/\!\widehat p  \nnb \\
   &&+ ( i S_{11} \widehat p^{[\alpha} \gamma^{\beta]} +
   2S_{12} \sigma^{\alpha\beta} )
   \,\,\rlap/\!\widehat p \,(v\cdot \epsilon^\ast)
   + (i S_{13} \gamma^{[\alpha} 
   \epsilon^{\ast\beta]} +  i S_{14} \gamma^{[\alpha} \epsilon^{\ast\beta]}\,
   \,\rlap/\!\widehat p )\, \nnb \\ 
   &&+ (iS_{15}\epsilon^{\ast[\alpha}\widehat{p}^{\beta]}
    + iS_{16}\epsilon^{\ast[\alpha}\widehat{p}^{\beta]}\,\rlap/\!\widehat p )
  +( i S_{17}\epsilon^{\ast[\alpha}v^{\beta]} + \,iS_{18}\epsilon^{\ast[\alpha}
    v^{\beta]} \,\rlap/\!\widehat p )\, \nnb \\
  &&+ ( i S_{19} \widehat p^{[\alpha} v^{\beta]}\,\rlap/\epsilon^\ast 
    + i\,S_{20} \widehat p^{[\alpha} v^{\beta]}\,\rlap/\epsilon^\ast\,
    \rlap/\!\widehat p )
   + ( i S_{21} \widehat p^{[\alpha} v^{\beta]} + i\,S_{22}\widehat p^{[
   \alpha} v^{\beta]} \,\rlap/\!\widehat p  ) (v\cdot \epsilon^\ast) \nnb \\
   &&+ ( i S_{23} \gamma^{[\alpha} v^{\beta]}\,\rlap/\epsilon^\ast
    + \,i S_{24} \gamma^{[\alpha} v^{\beta]}\,\rlap/\epsilon^\ast\,\rlap/\!\widehat p )
   + (i S_{25} \gamma^{[\alpha} v^{\beta]} + \, i S_{26} 
   \gamma^{[\alpha} v^{\beta]}\,\rlap/\!\widehat p )(v\cdot \epsilon^\ast)\,\Big]\,,
\eeq
with $[\alpha,\beta]$ being the anti-symmetric index.
 Note that $ig_sG^{\alpha\beta}
=[\,i D^\alpha, i D^\beta\,]$, and 
\begin{eqnarray}
-\bar h_v\,\Gamma_1\, D^{\alpha}\,D^{\beta}\, h_v = - \partial ^{\alpha} 
( \bar h_v \, \Gamma_1 \,D^{\beta} \, h_v ) \,+\,\bar 
h_v\,\overleftarrow{D^\alpha}\, 
\Gamma_1\,D^{\beta}\,h_v\,\,.
\end{eqnarray}
After the integration over $x$, the total divergence term can be neglected.
Consequently, using the equation of motion of heavy quark 
$i\, v\cdot D\, h_v\,=0\,,$ one finds that
\begin{eqnarray}
v_{\beta}{\rm Tr}\bigg\{ 
   \, \bar \Omega_G^{\alpha\beta} (v,\, p)\,\Gamma\,{1+\rlap/v\over 2}\,
   \Gamma_1\,{\cal{M}}(v) \bigg\} = 0\,.
\end{eqnarray}
This means $v_{\beta}\bar \Omega_G^{\alpha\beta}=0$ which yields,
\beq
&& S_5=S_{19},\,\, 2S_6=S_{19}-S_{23},\,\, S_7+S_{15}=S_{21}, \nnb \\
&& 2S_8=S_{17}+S_{21}-S_{25},\,\, S_9=S_{20},\,\, 2S_{10}=-S_{24},\,\, 
   S_{11}+S_{16}=S_{22}, \nnb \\
&& 2S_{12}=S_{18}+S_{22}-S_{26},\,\, S_{13}+S_{15}=-S_{17},\,\, 
   S_{14}+S_{16}=-S_{18}\,. 
\eeq
Therefore, all of the twelve universal functions $S_i\,(i=5\sim 16)$
are independent. 

Until now all the independent universal functions for heavy meson to light 
vector meson transitions are obtained, four at leading order, twenty-two at 
next-to-leading order.  Note that even to the order of $1/m_Q$, the form 
factors in $B\to\rho$ semileptonic decay and that in $B\to K^*$ rare decays 
are connected.  

By evaluating the traces, we get the relevant form factors to the order of 
$1/m_Q$ in terms of the universal functions:
\beq
\tilde g& =& -L_3 + \frac{1}{2m_Q}\Big(D_2-D_7-2D_9-D_{14}-S_3-2S_5
           +2S_9 -6S_{10}-2S_{14}+S_{15}-S_{16} \Big), \nnb \\
\tilde f& =& - (L_1 + L_3) + \frac{1}{2m_Q}\Big( -D_1-D_2+4D_3-D_7-\widehat 
           p^2 D_8-2D_9+D_{13}-D_{14} \nnb \\
&&\quad\quad\quad\quad\quad\quad\quad
  -S_1-S_3 -6S_6+(1-\widehat p^2)S_9 -6S_{10}-2S_{13}-2S_{14}
  +(\widehat{p}^2-1) S_{16} \Big), \nnb \\
\tilde a_2& =& L_2 + \frac{1}{2m_Q}\Big( D_4+2D_6+\widehat p^2 D_{11}+S_2-2S_7
  +6S_8 -2\widehat p^2 S_{11}+2S_{13}-\widehat{p}^2S_{16}\Big), \nnb \\
\tilde a_1& =& (L_3 - L_4) + \frac{1}{2m_Q}\Big( -D_2+D_5+D_7+2D_9-D_{10}
-2D_{11}+4D_{12}-D_{14} \nnb \\
&&\quad\quad\quad\quad\quad\quad\quad\quad
  S_3-S_4+2S_5+2S_7-2S_9+6S_{10}+2S_{11}-6S_{12}+S_{16}  \Big),
\eeq
and
\beq
\tilde g_v &=& L_1 +\frac{1}{2m_Q}\Big( D_1+2D_2-3\widehat p^2D_8+D_{13}
                +S_1-S_5+6S_6+5\widehat p^2S_9+2S_{13}+S_{14}-S_{15}+3S_{16}  
                \Big), \nnb \label{bb} \\
\tilde g_p &=& L_3 +\frac{1}{2m_Q}\Big( D_2+D_7-4D_9-D_{14}+S_3+3S_5
                +S_9 +6S_{10}+3S_{14}+S_{15}+S_{16}  \Big), \nnb \\
\tilde h   &=& L_4 +\frac{1}{2m_Q}\Big( D_5+D_{10}+2 D_{12}+S_4-S_5-2S_7
             -2S_{11} +6S_{12}+S_{14}+S_{15} \Big), \nnb \\
\tilde g_v' &=& L_1 +\frac{1}{2m_Q}\Big( D_1+2D_2-\widehat p^2D_8+D_{13}
                +S_1-2S_5+6S_6 +2\widehat p^2S_9+3S_{13}+S_{15}-
                \widehat{p}^2S_{16}   \Big), \nnb \\
\tilde g_p' &=& L_3 +\frac{1}{2m_Q}\Big( D_2+D_7+2D_8-4D_9-D_{14}+S_3+2S_5
                -2S_9 +6S_{10}+2S_{14} -S_{15}+S_{16}    \Big), \nnb \\
\tilde h'  &=& L_4 +\frac{1}{2m_Q}\Big( D_5+D_{10}+2 D_{12}+S_4-2S_7
             -2S_{11} +6S_{12} +2S_{14}-S_{15}    \Big) \label{ee} \,.
\eeq

In summary, within the HQET,  we have systematically analyzed the heavy to 
light vector meson transitions to the order of $1/m_Q$.  Besides the four 
universal functions at the leading order, there are twenty-two independent 
universal form factors at the order of $1/m_Q$.  Both the semileptonic decay 
$B\to\rho$ which is relevant to the $|V_{ub}|$ extraction, and the penguin 
induced decay $B\to K^*$ which is important to new physics discovering, depend 
on these form factors.  Once they are given, we can use them to calculate all 
kinds of decays involving such transitions to a good precision.

Some model-independent observations can be made.  Consider the decay 
$B\to\rho\,\ell\,\nu$, the decay rate will be largely simplified if we work at 
the zero recoil point of $\rho$ meson.  Only $|\tilde f|^2$ has non-vanishing 
contribution.  So we have
\beq
&&\left. \frac{{\rm d}\Gamma (B\to \rho\,\ell\,\bar\nu_\ell)}
  {{\rm d}(v\cdot p)}\,
  \right|_{v\cdot p \sim m_\rho}
  = {G_F^2\, |V_{ub}|^2\over24\,\pi^3}\sqrt{(v\cdot p)^2-m_\rho^2}
 \,\cdot \,\frac{Q(v\cdot p)-m_l^2}{Q(v\cdot p)} \nnb \\ 
&& \quad \times \bigg\{ \bigg[- (L_1 + L_3) + \frac{1}{2m_b}
    \Big( 2(D_1-D_8+D_{13})+(\bar\Lambda+v\cdot p)L_1+
    (3\bar\Lambda -v\cdot p \,
    \widehat p^2)L_3  \nnb \\
&&\quad\quad \quad -S_1-S_3 -6S_6 +(1-\widehat p^2)S_9  
-6S_{10} -2S_{13}-2S_{14} +(\widehat{p}^2-1)S_{16} \Big)\,\bigg]^2 \nnb \\
&&\quad\quad \times \bigg[\, 
  \frac{m_B^2}{m_\rho^2} \Big( \,(v\cdot p)- \frac{m_\rho^2}{m_B}\,\Big)^2
  \Big(1- \frac{m_l^4}{Q^2(v\cdot p)}\Big) -2\Big( Q(v\cdot p)-m_l^2\Big)\,
  \bigg]\,\bigg\}\,,
\eeq 
where $Q(v\cdot p)=m_B^2+m_\rho^2-2m_B(v\cdot p)$. Only  
thirteen universal functions are needed to determine the decay rate at zero 
recoil point.  We may also include the flavor changing neutral current decays, 
such as $B\to K^\ast \gamma$ and $B\to K^\ast \ell\,\bar\ell$ to get some 
information of the unknown universal functions appeared at the order of 
$1/m_Q$.  

The $1/m_Q$ corrections are more important for the decays $D\to K^*$ and 
$D\to\rho$.  The information of the $1/m_Q$ corrections can be drawn through 
the following way.  First, from the $B\to\rho$ decay, neglecting the $1/m_b$ 
effect, we can get certain result for the leading order heavy quark expansion 
through comparing with the experimental data.  Then input this knowledge to 
the $D\to K^*(\rho)$ decays, while keeping the $1/m_c$ corrections, the 
information of the $1/m_Q$ correction to the decay can be obtained with an 
uncertainty subject to $m_c/m_b\sim 30\%$.  

To obtain more detailed results of the decays, the knowledge about the 
universal form factors themselves are needed.  While the HQS simplifies the 
analysis, it does not predict the $v\cdot p$ dependence of the universal 
functions. This dependence must be determined separately by using 
nonperturbative techniques, such as QCD sum rules or lattice simulation, which 
are the next important steps to obtain quantitative results.

\begin{center}
{\large\bf Acknowledgments}\\[10mm]
\end{center}\par

One of us (C.L.) would like to thank Prof. H. S. Song for the helpful
discussions.  This work was supported in part by the National Natural Science 
Foundation of China and the BK21 Program of the Ministry of 
Education of Korea.

\end{document}